\documentclass[twocolumn,preprintnumbers,amsmath,amssymb]{revtex4}

\usepackage{epsfig}
\usepackage{amsmath}

\begin{document}

\title{Short note on magnetic impurities in SmFeAsO$_{1-x}$F$_x$ (x=0, 0.07) compounds revealed by zero-field $^{75}$As NMR}

\author{A. A. Sidorenko}
\email{sidorenko@fis.unipr.it}
\author{R. De Renzi}
\affiliation{Dipartimento di Fisica e Unit\`a CNISM, Universit\`a degli Studi di Parma, Vialle delle Scienze, 7A, 43100 Parma, Italy}
\author{A. Martinelli}
\author{A. Palenzona}
\affiliation{CNR-INFM-LAMIA Artificial and Innovative Materials Laboratory, Corso Perrone 24, 16152 Genova, Italy}
\date{\today}

\begin{abstract}
We have performed zero-field $^{75}$As nuclear magnetic resonance study of SmFeAsO$_{1-x}$F$_x$ (x=0, 0.07) polycrystals in a wide frequency range at various temperatures. $^{75}$As resonance line was found at around 265 MHz revealing the formation of the intermetallic FeAs clusters in the new layered superconductors. We have also demonstrated that NMR is a sensitive tool for probing the quality of these materials. 
\end{abstract}

\keywords{superconductivity, SmFeAsO, NMR}

\maketitle

The layered rare-earth metal oxypnictides ROFeAs (where R - rare-earth ions) have attracted a great attention after the discovery of the superconductivity in the iron-based LaFeAsO$_{1-x}$F$_x$.\cite{Kamihara} Replacement of La by other rare-earth ions leads to a large increase in T$_C$ from 26 K in LaFeAsO$_{1-x}$F$_x$ to T$_C>50$ K in RFeAsO$_{1-x}$F$_x$ with R = Nd, Pr, Sm, and Gd.\cite{Ren1,Liu,Ren2,Yang,Chen} However, \textit{Nowik and Felner}\cite{Nowik} using M\"{o}ssbauer spectroscopy have recently found that most RFeAsO$_{1-x}$F$_x$ compounds contain foreign magnetic Fe-As phases such as Fe$_2$As, FeAs$_2$ and FeAs, regardless the preparation method in amounts which might reach even 50\%. Although large amounts are easily detected by X-ray diffraction (XRD) quanteties below 5 - 10\% may escape this standard characterization and influency heavily magnetic properties as seen by macroscopic magnetization/susceptibility. In principle, this fact can lead to wrong conclusions on magnetic and related properties of the newly discovered superconductors.

In this work, we investigate SmFeAsO$_{1-x}$F$_x$ compounds with x=0 and 0.07 addressing specifically the issue of intermetallic magnetic FeAs clusters with the magnetic transition temperature 77 K\cite{Kulshreshtha} by means of zero-field $^{75}$As nuclear magnetic resonance (NMR). NMR spectra were collected with the home-built broadband fast-averaging NMR spectrometer HyReSpect\cite{Allodi} on a tuned probe circuit. The zero-field spectra were obtained in the frequency range 200-350 MHz by means of a standard optimized $\Theta-\tau-\Theta$ spin-echo pulse sequence, plotting point by point the amplitude at zero frequency shift of the Fast Fourier Transform of each echo as a function of transmission frequency. The NMR spectra are always corrected for NMR sensitivity, rescaling their amplitudes by $\omega^{2}$. Details on the samples preparation as well as on their characterization by means of Rietveld refinment of XRD data, scanning electron microscopy observation, transmission electron microscope analysis, resistivity and magnetization measurements can be found elsewhere.\cite{Martinelli} Note that, our samples do not show traces of FeAs by XRD. However, in \cite{Martinelli} a very small amount of ferromagnetic impurities was supposed from results of SQUID measurements.

\begin{figure}
\centerline{\psfig{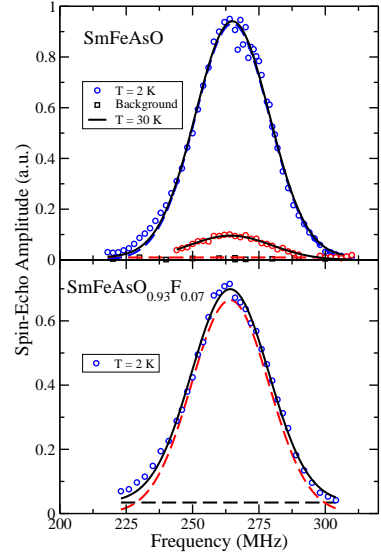}}
\caption{Zero-field $^{75}$As NMR measured in SmFeAsO$_{1-x}$F$_x$ compounds with x=0 and 0.07 at T = 2 and 30 K.}
\label{fig:spectra}
\end{figure}

The NMR frequency is very sensitive to the local environment of the nucleus. In the magnetic materials, such as FeAs, due to the collective electrons and the overlapping of wave functions of the adjacent atoms, the spin polarization is transferred from the magnetic ions (Fe in our case) to the nonmagnetic atoms (As) giving rise to the hyperfine fields of the Fermi-contact origin on their nuclei. Therefore, the NMR frequency of nonmagnetic As ions in zero magnetic field strongly depends on the magnetic and conducting state of the compounds. Figure \ref{fig:spectra}(a) shows the $^{75}$As NMR spectra of the SmFeAsO sample measured in zero external magnetic field at low temperature. We detected a quite broad Gaussian resonance line centered at $\sim265$ MHz with a full width at half maximum (FWHM) of 30 MHz. As might be seen in Figure \ref{fig:spectra}(b) very similar spectrum was found in the doped sample SmFeAsO$_{0.93}$F$_{0.07}$ as well. It should be mentioned that the position and FWHM of these lines do not depend on the doping level, that is, the resonance lines are the very same in both normal (undoped) and superconducting (doped) phases. In addition, by applying an external magnetic field the resonance line is shifted towards high frequencies, indicating a positive hyperfine field on nuclei, as expected for the As ions.\cite{Freeman} The existence of the resonance in zero external magnetic field and low applied radio-frequency power indicate that the NMR signal is due to the enhancement of the radio-frequency field by domain wall motion\cite{Turov,Portis} revealing the formation of the magnetic FeAs clusters (maybe even of the nanoscopic dimension) in SmFeAsO$_{1-x}$F$_x$ compounds and, therefore, confirming the results of M\"{o}ssbauer spectroscopy.\cite{Nowik}

\bibliography{Text}

\end{document}